\documentclass[reprint,superscriptaddress,amsmath, amssymb,aps,prd,notitlepage,longbibliography,floatfix,nofootinbib,twocolumn]{revtex4-1}

\usepackage{amsmath}
\usepackage{amssymb}
\usepackage{slashed}
\allowdisplaybreaks[4]
\usepackage{cancel}
\usepackage{extarrows}
\usepackage{tensor}     % manipulate tensors
\usepackage{float}
\usepackage[caption = false]{subfig} % combine multiple figures
\usepackage[final]{graphicx}   % include figures
\usepackage[
colorlinks=true,        % color link
citecolor=blue,         % cite color
linkcolor=blue,         % link color
urlcolor=blue           % url color
]{hyperref}             % create hyperlinks
\usepackage{bm}         % \bm{<text>} Bold math symbols
\usepackage{xcolor}     % textcolor
\usepackage{lipsum}
\usepackage{color}      % \textcolor{declared-color}{text}
\usepackage[utf8]{inputenc} % accented characters for .bib file using XeLaTex
\usepackage[section]{placeins} % figures processing
\usepackage{appendix}
\usepackage{units}
\usepackage[capitalise]{cleveref}
\usepackage{units}
\newcommand{\nc}{\newcommand*}

\makeatletter
\def\@seccntformat#1{\csname the#1\endcsname.\quad}
\makeatother
\nc{\xbar}{\bar{x}}
\nc{\rhoeq}{\rho_{\mathrm{eq}}}
\nc{\zeq}{z_{\mathrm{eq}}}
\nc{\tla}{\tilde{\lambda}}
\nc{\bt}{\beta}
\nc{\dt}{\delta}
\nc{\Dt}{\Delta}
\nc{\vj}{\vec{j}}
\nc{\vl}{\vec{l}}
\nc{\hx}{\hat{x}}
\nc{\hy}{\hat{y}}
\nc{\bj}{\bm{j}}
\nc{\mJ}{\mathcal{J}}
\nc{\mP}{\mathcal{P}}
\nc{\Msun}{M_\odot}
\nc{\av}[1]{\langle #1 \rangle}
\nc{\eq}[1]{Eq.~\eqref{#1}}
\nc{\al}{\alpha}
\nc{\Xstar}{X_{\ast}}
\nc{\fpbh}{f_{\mathrm{pbh}}}
\nc{\vth}{\vec{\theta}}
\nc{\vla}{\vec{\lambda}}
\nc{\vd}{\vec{d}}
\nc{\Mmin}{M_{\mathrm{min}}}
\nc{\rmd}{\mathrm{d}}
\nc{\mmin}{{m_{\mathrm{min}}}}
\nc{\mmax}{{m_{\mathrm{max}}}}
\nc{\mR}{\mathcal{R}}
\nc{\tmR}{\tilde{\mathcal{R}}}
\nc{\s}{\sigma}
\nc{\ogw}{\Omega_{\mathrm{GW}}}
\nc{\addref}{[\textcolor{red}{add ref}] }
\nc{\Om}{\Omega}
\nc{\gm}{\gamma}
\nc{\Gm}{\Gamma}
\nc{\gpcyr}{\mathrm{Gpc}^{-3}\,\mathrm{yr}^{-1}}
\nc{\Eq}[1]{Eq.~\eqref{#1}}
\nc{\Fig}[1]{Fig.~\ref{#1}}
\nc{\Table}[1]{Table~\ref{#1}}
\nc{\lvc}{LIGO/Virgo} % LIGO-VIRGO collaboration
\nc{\Sec}[1]{Sec.~\ref{#1}}
\nc{\eg}{\textit{e.g.~}}

\nc{\sovast}{Soviet Ast.}
\begin{document}

\title{Gravitational Wave Tails and Transient Behaviors of Quantum-Corrected Black Holes}

%%%%%%%%%%%%%%%%%%%%%%%%%%%%%%%%%%%% author
%%%%%%%%%%%%%%%%%%%%%%%%%%%%%%%%%%%%
\author{Rong-Zhen Guo}
\email{guorongzhen@ucas.ac.cn}
\affiliation{School of Fundamental Physics and Mathematical Sciences, Hangzhou Institute for Advanced Study, UCAS, Hangzhou 310024, China}
\affiliation{School of Physical Sciences, 
    University of Chinese Academy of Sciences, 
    No. 19A Yuquan Road, Beijing 100049, China}
%\affiliation{CAS Key Laboratory of Theoretical Physics, Institute of Theoretical Physics, Chinese Academy of Sciences,Beijing 100190, China}

%%%%%%%%%%%%%%%%%%%%%%%%%%%%%%%%%%%%
\author{Qing-Guo Huang}
\email{corresponding author: huangqg@itp.ac.cn}
\affiliation{School of Fundamental Physics and Mathematical Sciences, Hangzhou Institute for Advanced Study, UCAS, Hangzhou 310024, China}
\affiliation{School of Physical Sciences, 
    University of Chinese Academy of Sciences, 
    No. 19A Yuquan Road, Beijing 100049, China}
\affiliation{Institute of Theoretical Physics, Chinese Academy of Sciences,Beijing 100190, China}

%%%%%%%%%%%%%%%%%%%%%%%%%%%%%%%%%%%%%%%%%%%%%%%%
%%%%%%%%%%%%%%%%%%%%%%%%%%%%%%%%%%%%%%%%%%%%%%%%%%%%%%

\begin{abstract}
Gravitational wave astronomy plays a pivotal role in testing the dynamics of gravity in strong-field regimes and probing the nature of black holes. Motivated by recent studies on late-time tails in gravitational waves, we examine the gravitational wave tails of black holes incorporating quantum corrections within the framework of effective Loop Quantum Gravity. Our findings indicate that both the amplitudes and the intermediate behavior of these tails are influenced by quantum corrections. We demonstrate that the amplitude and transient characteristics of the tail are sensitive to the specific details of the black hole's dynamics.

\end{abstract}
% \flushbottom
% \pacs{}
\maketitle

\section{Introduction}

The development of gravitational-wave (GW) astronomy has revolutionized our ability to probe the properties of compact objects \cite{TheLIGOScientific:2014jea, TheVirgo:2014hva, TheLIGOScientific:2016htt, TheLIGOScientific:2016src, O1_release, Will:2014kxa, Zurek:2022xzl}. As the next generation of both space-based and ground-based detectors is deployed, we expect more stringent tests of gravitational theory. In particular, these advancements will allow for novel investigations into potential deviations from the predictions of Einstein's general relativity (GR), which has remained the most successful theory of gravity to date. These efforts are essential for further refining GR, as evidenced by recent studies \cite{Berti:2018cxi, Reitze:2019iox, Baker:2019nia, Perkins:2020tra, TianQin:2015yph, Hu:2017mde, LIGOScientific:2025rid}.

Quantum effects in black holes (BHs) may leave subtle signatures on GWs emitted during binary BH mergers \cite{Cardoso:2019rvt, Barack:2018yly}. These effects are expected to influence the background geometry. On one hand, they may preserve the event horizon, while simultaneously modifying the geometric structure of the background. This scenario applies to regular BHs \cite{Lan:2023cvz, Chamseddine:2016ktu, Nicolini:2005vd}, higher-dimensional BHs \cite{Myers:1986un, Emparan:2008eg, Emparan:2009vd}, and phenomenological BH models predicted by quantum gravity theories \cite{Zhang:2024khj, Lewandowski:2022zce, Garfinkle:1990qj, Youm:1997hw, Chow:2014cca}.

On the other hand, many theories predict the existence of objects that do not possess event horizons, often called BH mimickers or exotic compact objects (ECOs) \cite{Bambi:2025wjx}. These include wormholes \cite{Morris:1988cz, Lobo:2009ip, Huang:2019arj}, firewalls \cite{Almheiri:2012rt}, fuzzballs \cite{Mathur:2005zp, Skenderis:2008qn}, gravatsars \cite{Rahaman:2012xx, Mazur:2004fk, Mazur:2001fv}, and gravitational solitons in supergravity \cite{Dima:2025tjz, Chakraborty:2025ger}, as well as models such as BH area quantization \cite{Bekenstein:1995ju}. Within BH perturbation theory, these different models either modify the background, altering the effective potential, or introduce boundary conditions that differ from those of classical BHs, thereby influencing GW propagation after binary BH mergers.

For example, in BH spectroscopy, these modifications can shift the quasinormal modes (QNMs) of BHs, enabling the detection of their nature through the search for higher overtones \cite{Berti:2005ys, Dreyer:2003bv, Bekenstein:1995ju, Cabero:2019zyt, Giesler:2019uxc, Berti:2025hly}. However, extracting these QNMs from GW signals remains a complex challenge due to spectral instability \cite{Jaramillo:2020tuu, Baibhav:2023clw, Boyanov:2022ark, Crescimbeni:2025ytx}. Furthermore, for ECOs, modifications near the horizons may generate GW echoes. Numerous studies have shown that echoes can arise in a variety of models \cite{Bueno:2017hyj, Biswas:2022wah, Rosato:2025lxb, Cardoso:2017cqb, Cardoso:2019apo, Fang:2021iyf}, with significant progress made in constructing echo waveforms from numerical relativity \cite{Ma:2022xmp}. However, no conclusive evidence of echo signals has been found in real observational data to date. Additionally, superradiant instability and its observable effects provide another avenue for studying quantum effects in BHs \cite{Guo:2021xao, Guo:2023mel, Zhou:2023sps}.

Current research primarily focuses on the early stages of the ringdown phase, which is well described by a superposition of QNMs. In fact, the post-merger GW signal also contains additional low-frequency components, such as GW tails. Tails arise from the backscattering of GWs in a non-flat spacetime background \cite{Price:1971fb, Gundlach:1993tp, Ching:1995tj}. Mathematically, they can be interpreted as originating from a branch cut in the Green function within perturbation theory, leading to their characteristic polynomial decay behavior. In \cite{Zenginoglu:2012us}, the authors demonstrated that this decay behavior is ultimately reached through the superposition of several other power laws at intermediate times. Several studies have already investigated the intermediate behavior \cite{Cardoso:2024jme, DeAmicis:2024eoy, Islam:2024vro, Ma:2024hzq, DeAmicis:2024not, Alnasheet:2025mtr}. Notably, in \cite{Alnasheet:2025mtr}, the authors, through numerical computations of a massless scalar field in a metric background modified by environmental effects, found that these effects could imprint on the late-time tails by influencing the intermediate behavior, thereby affecting the amplitudes of the tails. These results offer valuable insights for studying quantum effects on BHs. In phenomenological models of BHs with quantum corrections, many examples show that the background geometries are directly modified. This suggests that both the late-time tails and the transition from QNM-dominated to tail-dominated behavior may be influenced by these quantum corrections.

In this study, we calculate the GWs generated by the perturbation of quantum-corrected BHs within the framework of effective Loop Quantum Gravity (LQG), focusing specifically on the tails and the transition from a QNM-dominated regime to a tail-dominated phase. Our computational results show that quantum effects, similar to environmental influences and other factors, can imprint on late-time tails. While these effects may not produce observable signatures in GW astronomy, our findings confirm that the amplitudes of the tails are sensitive to the dynamical characteristics of the GW source. This underscores the importance of caution when constructing precise GW waveforms in future studies. We adopt geometrized units with $G=c=1$.

\medskip

\section{Setup} 

We consider a quantum black hole model derived from effective LQG \cite{Lewandowski:2022zce}, which can be viewed as a quantum extension of the Oppenheimer–Snyder collapse scenario. In the classical setting, the interior of a homogeneous, pressureless dust ball is described by a Friedmann–Lemaître–Robertson–Walker (FLRW) geometry. Quantum effects are incorporated through Loop Quantum Cosmology, leading to a modified effective spacetime description. The resulting line element takes the form
\begin{equation}
\mathrm{d}s^2 = -F(r),\mathrm{d}t^2 + F(r)^{-1},\mathrm{d}r^2 + r^2,\mathrm{d}\vartheta^2 + r^2 \sin^2\vartheta,\mathrm{d}\varphi^2,
\end{equation}
with
\begin{equation}
F(r) = 1 - \frac{2M}{r} + 27 \left( \frac{\alpha M}{2r} \right)^4,
\end{equation}
where $\alpha$ parametrizes the strength of quantum corrections.
Relative to the formulation in \cite{Lewandowski:2022zce}, we have rewritten the metric such that $\alpha$ is dimensionless. The classical Schwarzschild solution is recovered in the limit $\alpha = 0$. When $\alpha = 1$, the spacetime reaches an extremal configuration, in which the outer event horizon coincides with the inner Cauchy horizon.

GWs emitted from this BH can be described by a modified Teukolsky formalism \cite{Li:2022pcy,Guo:2024bqe}. Within this framework, gravitational radiation is encoded in the leading-order perturbation of the Weyl scalar $\Psi_4^{(1)}$ in the Newman--Penrose (NP) formalism \cite{Teukolsky:1973ha,Press:1973zz,Teukolsky:1974yv,Newman:1961qr}. Following \cite{Guo:2024bqe}, $\Psi_4^{(1)}$ satisfies
\begin{widetext}
\begin{equation}
\left[(\Delta + 3\gamma - \bar{\gamma} + 4\mu + \bar{\mu})(D + 4\epsilon - \rho)
- (\bar{\delta} - \bar{\tau} + \bar{\beta} + 3\alpha + 4\pi)(\delta - \tau + 4\beta)
- 3\Psi_2 + 2\Phi_{11}\right] \Psi_4^{(1)}
= 4\varpi T_4,
\label{Equation}
\end{equation}
\end{widetext}
where the superscript indicates that $\Psi_4^{(1)}$ denotes the linear perturbation, distinguishing it from background quantities. All symbols follow standard NP conventions (see Appendix~\ref{NP}), and $\varpi$ denotes the mathematical constant $\pi$, introduced to avoid confusion with similarly labeled NP quantities. Since the exterior spacetime dynamics coincide with those of general relativity, the source term $T_4$ vanishes in the homogeneous case.

To obtain accurate and stable numerical simulations of GWs, we adopt the approach in \cite{Ripley:2020xby}, which is based on choosing a null tetrad that remains regular at future null infinity. The calculations are carried out in horizon-penetrating, hyperboloidally compactified (HPHC) coordinates $\{T, R, \vartheta, \varphi\}$, which ensure that the resulting equations automatically satisfy the appropriate boundary conditions. The transformation between the HPHC coordinates $\{T, R, \vartheta, \varphi\}$ and the standard coordinates $\{t, r, \vartheta, \varphi\}$ is given by
\begin{equation}
\begin{aligned}
R(r) & = \frac{L^2}{r}, \\
T(v, r) & = v - r - 4M \ln r ,
\end{aligned}
\end{equation}
where $L$ is a constant length scale associated with the compactification of the radial coordinate. The coordinate $v$ is closely related to the ingoing Eddington--Finkelstein null coordinate in the Schwarzschild spacetime and is defined as
\begin{equation}
dv \equiv dt + dr_* - dr .
\end{equation}
Here, $r_*$ denotes the tortoise coordinate, defined by
\begin{equation}
\frac{dr_*}{dr} = F(r)^{-1}.
\end{equation}
The definition of $v$ differs from the standard choice $v = t + r_*$ and is introduced to ensure that the NP scalars remain finite near the BH horizon, thereby improving numerical stability.

Additionally, the null tetrad is selected as:

\begin{equation}
\begin{aligned}
l^{\mu} = & \left\{\frac{64L^6M^2R^2 - 27a^4L^2M^4R^4 - 54a^4M^5R^5}{16L^{10}}, \right. \\
& \left. -\frac{R^2 \left(16L^8 - 32L^6MR + 27a^4M^4R^4 \right)}{32L^{10}}, 0, 0 \right\}; \\
n^{\mu} = & \left\{ 2 + \frac{4MR}{L^2}, \frac{R^2}{L^2}, 0, 0 \right\}; \\
m^{\mu} = & \left\{ 0, 0, -\frac{R}{\sqrt{2}L^2}, -\frac{iR \operatorname{Csc}\theta}{\sqrt{2}L^2} \right\}; \\
\bar{m}^{\mu} = & \left\{ 0, 0, -\frac{R}{\sqrt{2}L^2}, \frac{iR \operatorname{Csc}\theta}{\sqrt{2}L^2} \right\}.
\end{aligned}
\end{equation}

We expand $\Psi_4^{(1)}$ as:
\begin{equation}
\Psi_4^{(1)}(T, R, \theta, \varphi) = R \sum_{\ell=0}^{\infty} \sum_{m=-\ell}^{\ell} \Psi_{\ell m}(T, R) {}_{-2}Y_{\ell m}(\vartheta, \varphi),
\end{equation}
where ${}_{-2}Y_{\ell m}$ are the standard spherical harmonics, $\ell$ is the orbital number, and $m$ is the magnetic quantum number.

Substituting this ansatz into Eq. (\ref{Equation}), we obtain:
\begin{equation}
\left(C_{TT} \partial_T^2 + C_{TR} \partial_T \partial_R + C_{RR} \partial_R^2 + C_T \partial_T + C_R \partial_R + C\right) \Psi_{\ell m} = 0, \label{neq}
\end{equation}
where
\begin{equation}
\begin{aligned}
C_{TT} &= \frac{M^2 \left(L^2 + 2MR\right) C_{Aux}}{4L^8}, \\
C_{Aux} &=\left(64L^6 - 27a^4L^2M^2R^2 - 54a^4M^3R^3\right),\\
C_{TR} &= -\frac{8L^{10} - 64L^6 M^2 R^2 + 27a^4L^2 M^4 R^4 + 54a^4 M^5 R^5}{4L^8}, \\
C_{RR} &= -R^2 + \frac{2MR^3}{L^2} - \frac{27a^4 M^4 R^6}{16L^8}, \\
C_T &= 8M - \frac{27a^4 M^4 R^3}{L^6} - \frac{243a^4 M^5 R^4}{4L^8}, \\
C_R &= 2R + \frac{2MR^2}{L^2} - \frac{135a^4 M^4 R^5}{8L^8}, \\
C &= -2 + \ell + \ell^2 - \frac{2MR}{L^2} - \frac{783a^4 M^4 R^4}{16L^8}.
\end{aligned}
\end{equation}

Several methods have been developed for stable long-time numerical simulations of GW initial value problems \cite{PanossoMacedo:2023qzp, Zenginoglu:2007jw, Zenginoglu:2011zz, PanossoMacedo:2014dnr, He:2025ydh, OBoyle:2022yhp, Markakis:2023pfh}. In this work, we adopt the time-symmetric integration scheme proposed in \cite{He:2025ydh, OBoyle:2022yhp, Markakis:2023pfh}. Unlike standard Runge--Kutta methods, this scheme is free from the Courant stability constraint and yields smaller truncation errors. It also preserves Noether charges over long time intervals, making it particularly suitable for long-term numerical evolutions in BH perturbation theory.

The radial coordinate $R$ is discretized using a Chebyshev pseudospectral method. The $N+1$ Chebyshev--Gauss--Lobatto collocation points are
\begin{equation}
R_i^{\mathrm{Cheb}} = \frac{R_{\mathrm{H}}}{2}
\left( 1 + \cos \frac{i\pi}{N} \right),
\quad i = 0, \ldots, N ,
\end{equation}
where $R_{\mathrm{H}}$ denotes the location of the event horizon in HPHC coordinates. Spatial derivatives in the $R$ direction are approximated using the corresponding differentiation matrix $\boldsymbol{D}$. To reduce Eq.~(\ref{neq}) to a first-order system, we introduce the auxiliary variable \cite{He:2025ydh, Ripley:2020xby}
\begin{equation}
\Pi_{\ell m}
=
\left(
C_{TT} \partial_T
+ C_{TR} \partial_R
+ C_T
\right)
\Psi_{\ell m}.
\end{equation}

This formulation allows the evolution equation to be written as
\begin{equation}
\frac{d\mathbf{u}}{dT} = \mathbf{L} \mathbf{u},
\end{equation}
where
\begin{equation}
\mathbf{u} = 
\begin{pmatrix}
\Psi_{\ell m} \\
\Pi_{\ell m}
\end{pmatrix},
\end{equation}
and
\begin{equation}
\mathbf{L} = 
\begin{pmatrix}
- \boldsymbol{L}_{\psi \psi} & \operatorname{diag}\left[ \frac{1}{C_{TT}} \right] \\
- \boldsymbol{L}_{\Pi \psi} & \mathbf{0}
\end{pmatrix},
\end{equation}
with
\begin{equation}
\boldsymbol{L}_{\psi \psi} = \operatorname{diag}\left[ \frac{C_{TR}}{C_{TT}} \right] \boldsymbol{D} 
+ \operatorname{diag}\left[ \frac{C_T}{C_{TT}} \right],
\end{equation}
\begin{equation}
\boldsymbol{L}_{\Pi \psi} = \operatorname{diag}\left[ C_{RR} \right] \boldsymbol{D}^2 
+ \operatorname{diag}\left[ C_R \right] \boldsymbol{D} 
+ \operatorname{diag}\left[ C \right].
\end{equation}

Time-symmetric evolution is implemented using the Hermite integration rule:
\begin{equation}
\begin{aligned}
\mathbf{u}_{v+1} &= 
\left[\mathbf{I} - \frac{\mathbf{L} \Delta T}{2} \left(\mathbf{I} - \frac{\mathbf{L} \Delta T}{6} \right)\right]^{-1} 
\left[\mathbf{I} + \frac{\mathbf{L} \Delta T}{2} \left(\mathbf{I} + \frac{\mathbf{L} \Delta T}{6} \right)\right] \mathbf{u}_v \\
&= \mathbf{u}_v + \left[\mathbf{I} - \frac{\mathbf{L} \Delta T}{2} \left(\mathbf{I} - \frac{\mathbf{L} \Delta T}{6} \right)\right]^{-1} \mathbf{L} \Delta T \mathbf{u}_v,
\end{aligned}
\end{equation}
where $\mathbf{u}_v$ denotes the value at $T = T_v$, and $\Delta T$ is the time step. Achieving stable numerical results requires high-precision floating-point arithmetic.

The symmetric initial data are chosen as
\begin{equation}
\Psi_{\ell m}(T=0, R) = k \exp \left[-\left(\frac{R - R_0}{w}\right)^2 \right] + b,
\end{equation}
\begin{equation}
\partial_T \Psi_{\ell m}(T=0, R) = 0,
\end{equation}
where $R_0$ and $w$ are the initial location and width of the Gaussian wave packet. In subsequent simulations, the overall amplitude is set to $k=1$. The parameter $b$ allows modeling of non-compact initial data: $b = 1$ corresponds to the non-compact case, while $b = 0$ represents compact initial data.

\medskip

\section{Results}

\begin{figure*}[htbp]
    \centering
    % 第一行
    \begin{minipage}{0.45\textwidth}
        \centering
        \includegraphics[width=\linewidth]{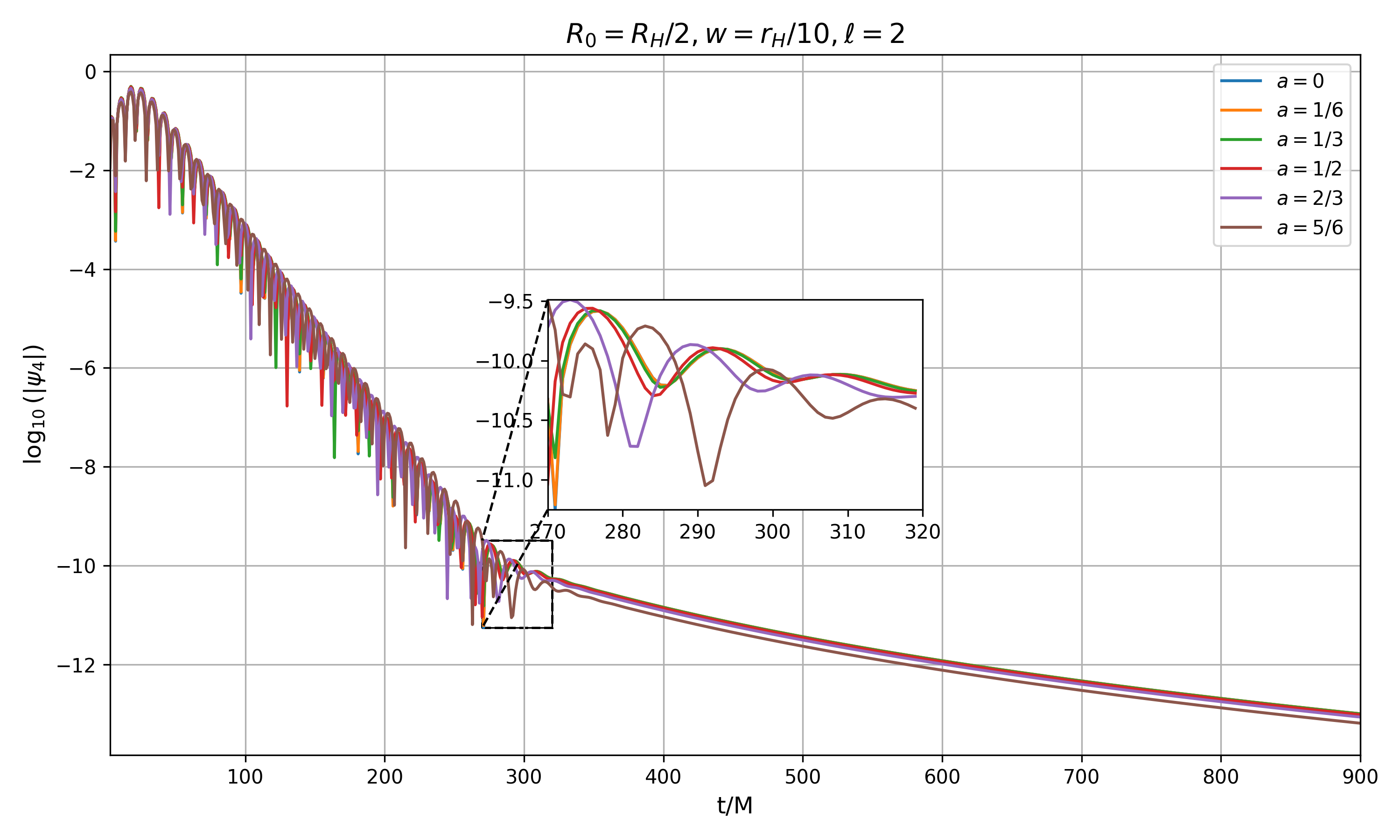}
    \end{minipage}%
    \hspace{0.05\textwidth}
    \begin{minipage}{0.45\textwidth}
        \centering
        \includegraphics[width=\linewidth]{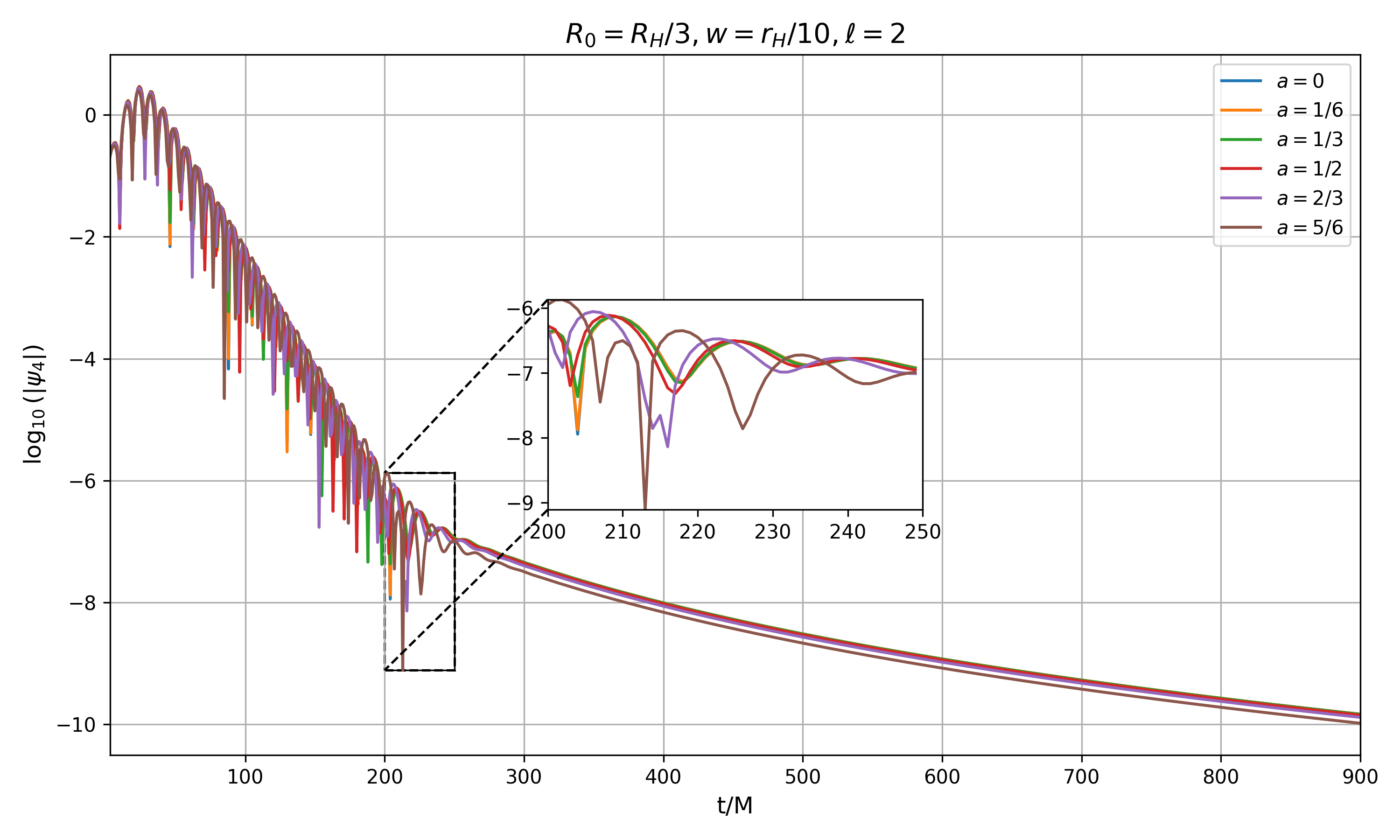}
    \end{minipage}

    \vspace{0.5cm}

    % 第二行
    \begin{minipage}{0.45\textwidth}
        \centering
        \includegraphics[width=\linewidth]{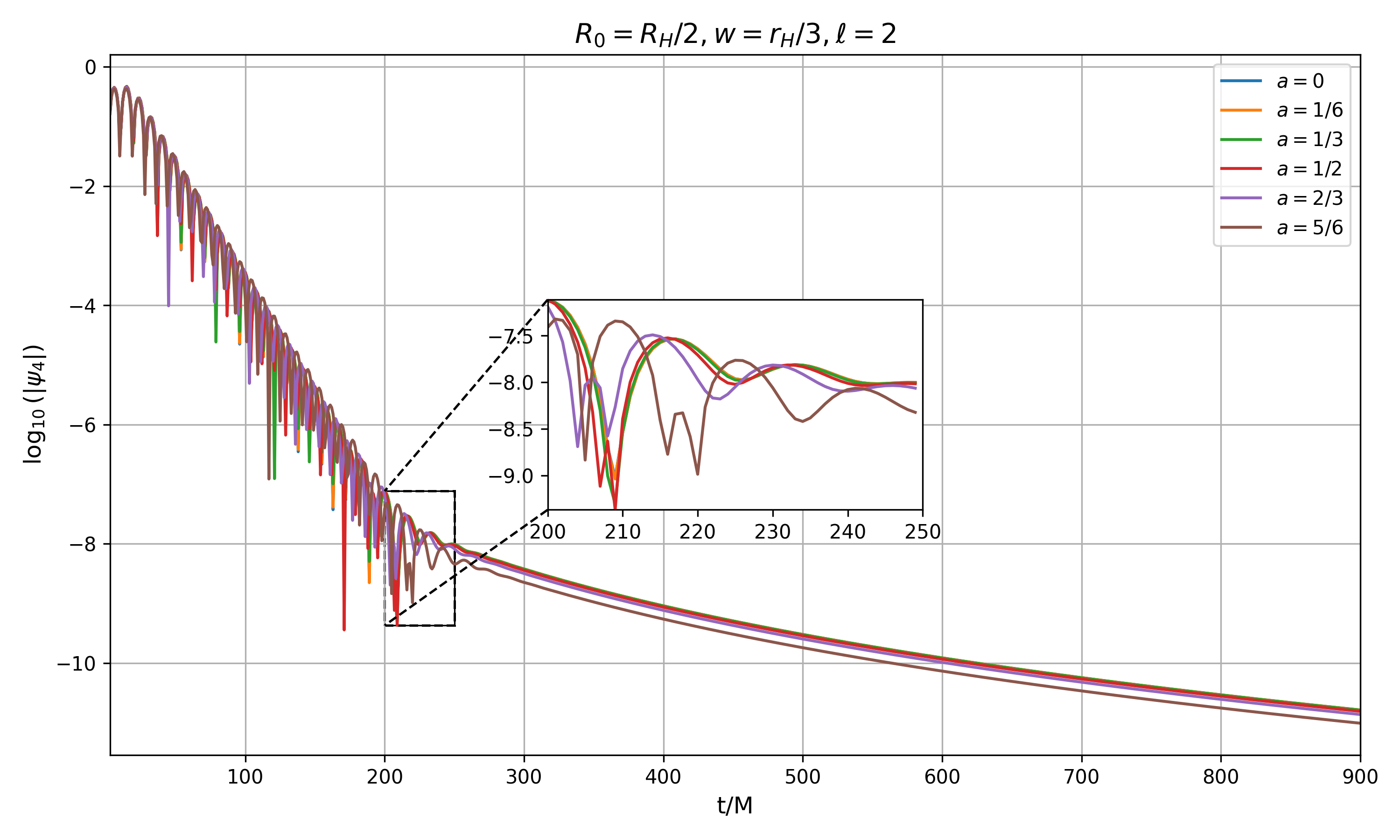}
    \end{minipage}%
    \hspace{0.05\textwidth}
    \begin{minipage}{0.45\textwidth}
        \centering
        \includegraphics[width=\linewidth]{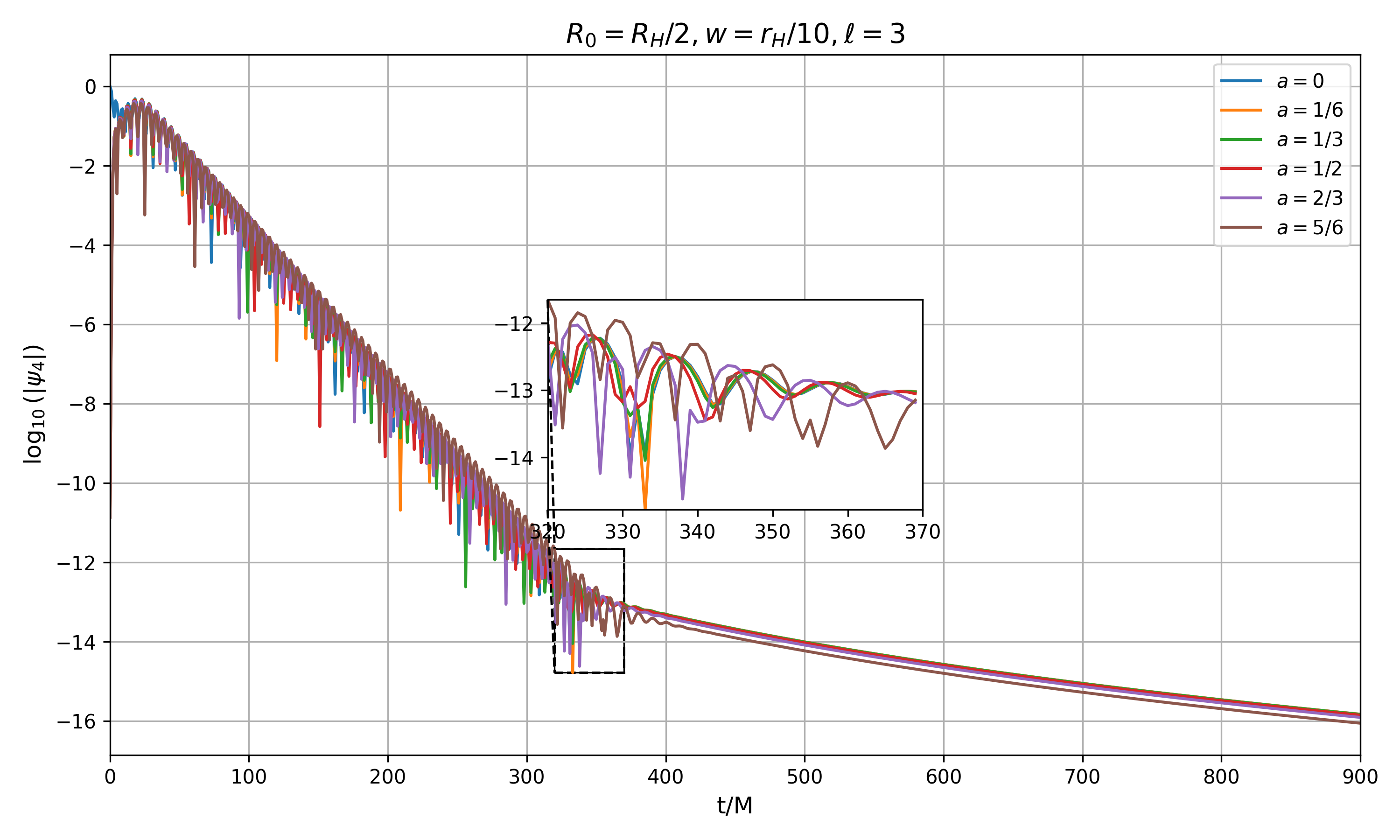}
    \end{minipage}

    \vspace{0.5cm}

    % 第三行（单张跨行图片）
    \begin{minipage}{0.45\textwidth} % 宽度设置为0.9，跨行
        \centering
        \includegraphics[width=\linewidth]{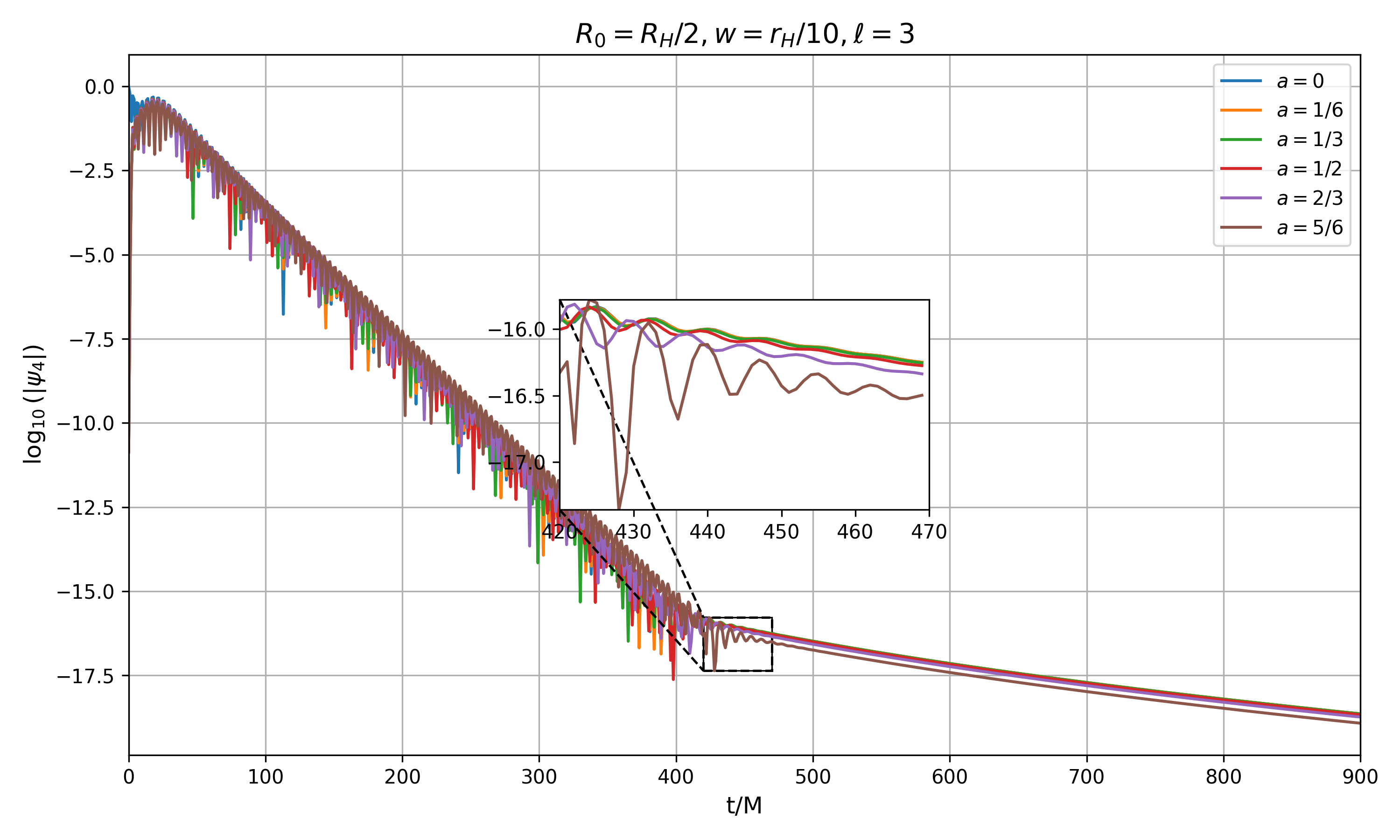}
    \end{minipage}

    \caption{Late-time tails for compact Gaussian initial data. Results are shown for varying quantum modification parameter $\alpha$, the center of the Gaussian wave packet $R_0$, and width $w_0$. In the late-time tail regime, the power-law decay is consistent with that in the classical GR case. However, the amplitude of the tail and the transition from the QNM-dominated phase to the tail-dominated phase are sensitive to $\alpha$.}
    \label{1}
\end{figure*}

\begin{figure*}[htbp]
    \centering
    % 第一行
    \begin{minipage}{0.45\textwidth}
        \centering
        \includegraphics[width=\linewidth]{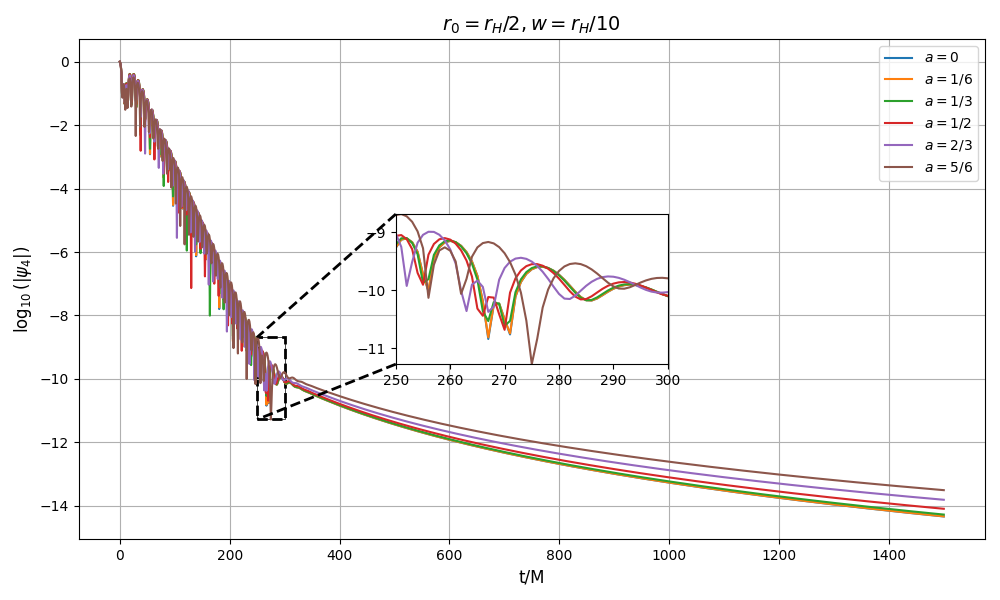}
    \end{minipage}%
    \hspace{0.05\textwidth}
    \begin{minipage}{0.45\textwidth}
        \centering
        \includegraphics[width=\linewidth]{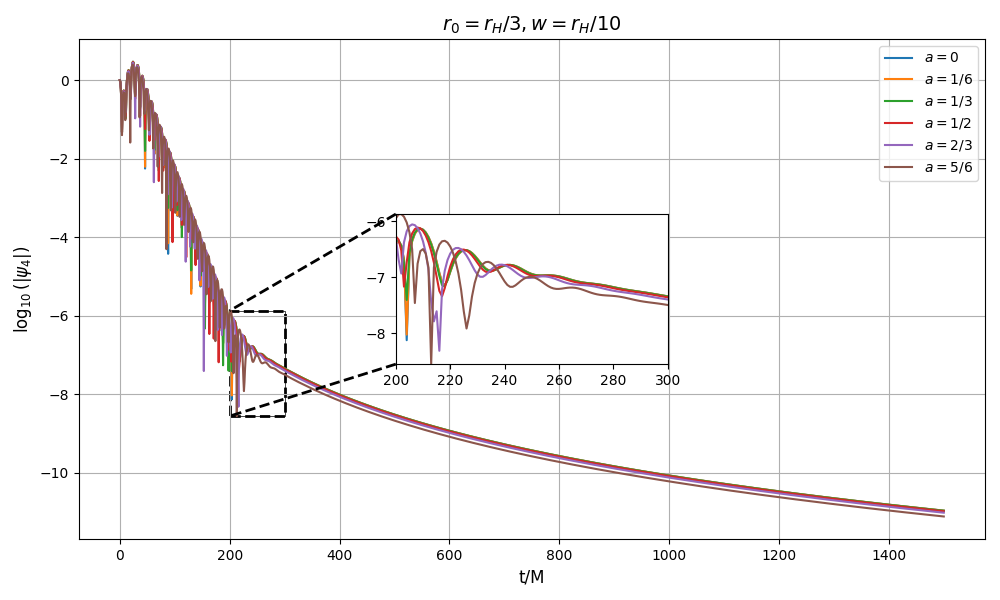}
    \end{minipage}

    \vspace{0.5cm}

    % 第二行
    \begin{minipage}{0.45\textwidth}
        \centering
        \includegraphics[width=\linewidth]{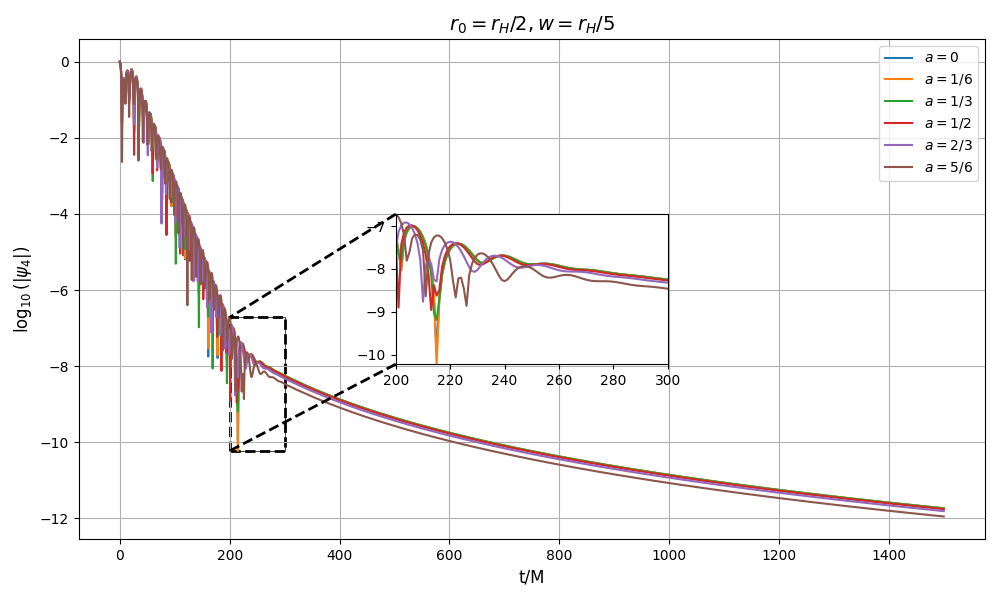}
    \end{minipage}%
    \hspace{0.05\textwidth}
    \begin{minipage}{0.45\textwidth}
        \centering
        \includegraphics[width=\linewidth]{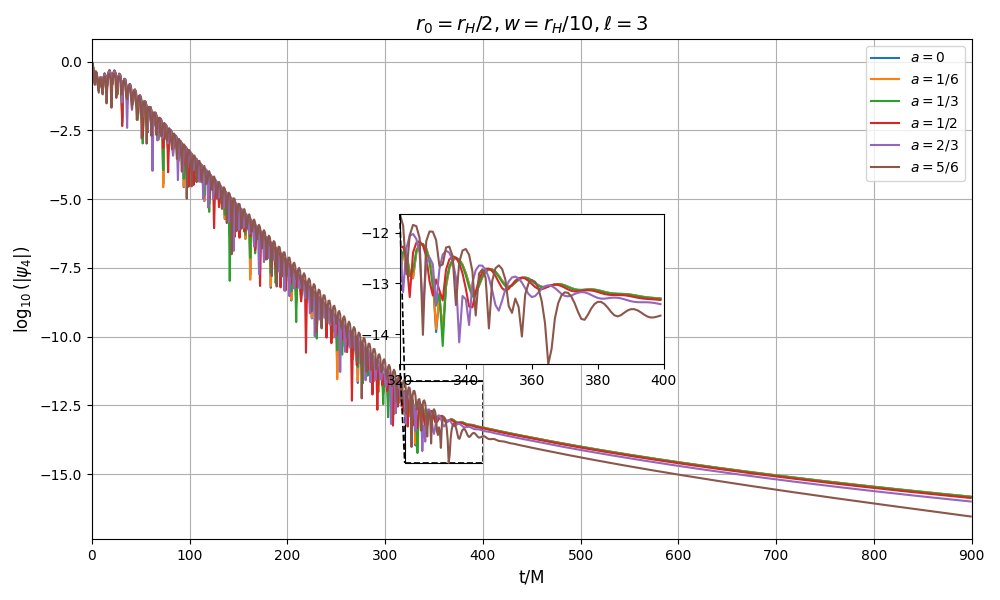}
    \end{minipage}

    \vspace{0.5cm}

    % 第三行（单张跨行图片）
    \begin{minipage}{0.45\textwidth} % 宽度设置为0.9，跨行
        \centering
        \includegraphics[width=\linewidth]{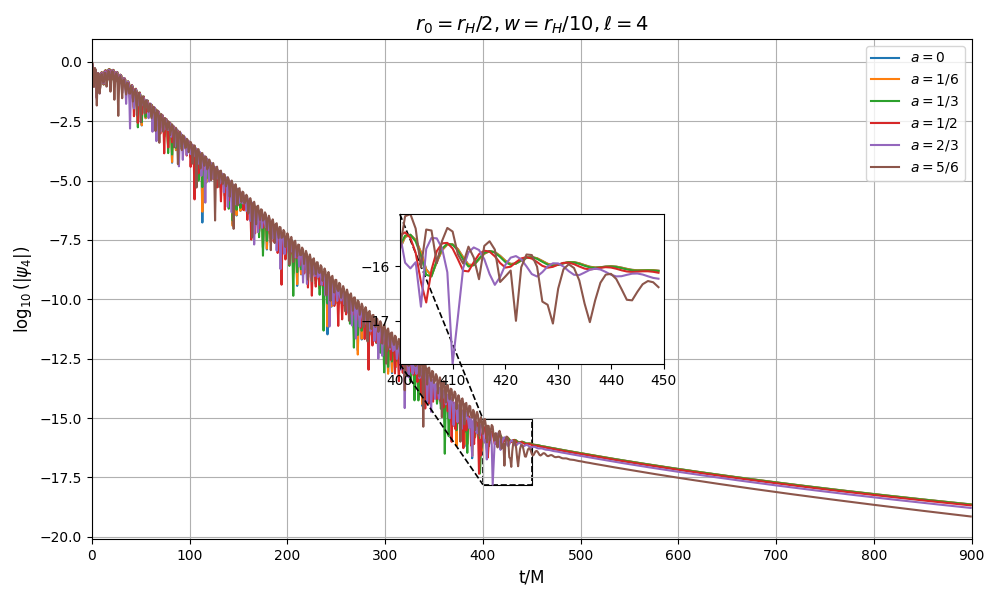}
    \end{minipage}

    \caption{Late-time tails for non-compact Gaussian initial data. Results are shown for varying quantum modification parameter $\alpha$, the center of the Gaussian wave packet $R_0$, and width $w_0$. Compared to compact initial data, non-compact initial data produces a relatively larger impact on the amplitude under the same parameter settings, with differences reaching up to an order of magnitude.}
    \label{2}
\end{figure*}

Our results are presented in Figs.~\ref{1} and \ref{2}, which summarize the outcomes of numerical evolutions conducted for initial data with varying locations, widths, multipoles, and the quantum modification parameter $\alpha$.

It is clear from the results that the late-time behavior in all cases aligns with classical GR predictions \cite{Gundlach:1993tp, Barack:1998bw}. This is expected, as the metric modification relative to GR is proportional to $r^{-4}$, meaning the effective potential retains the same asymptotic form as in GR \cite{Rosato:2025rtr}. However, we observe that the late-time amplitude is influenced by the quantum correction parameter $\alpha$.

The strength of this modification varies depending on factors such as the properties of the initial conditions. In general, simulations show that introducing $\alpha$ tends to reduce the amplitude of the late-time tails. The most significant amplitude change occurs in the case of non-compact initial conditions, where the amplitude can vary by an order of magnitude, as seen in the top-left and top-right images of the second row in Fig.~\ref{2}. In these cases, the quantum modification is comparable to the environmental effects discussed in \cite{Alnasheet:2025mtr}.

Another notable feature is the transient effects observed in the simulations. By varying the initial conditions, the amplitude of the late-time tail and the intermediate transition phase can yield different results, even for the same \((\ell, m)\) components. These effects are more pronounced in the case of non-compact initial conditions, while they have a lesser impact for compact conditions. A clear example of this can be seen by comparing the upper-left corners of Fig.~\ref{1} and Fig.~\ref{2}. Changes in initial condition compactness can cause the quantum modification to transition from a significant effect ($O(1)$) to nearly negligible. Additionally, for larger \(\ell\) values, increasing the quantum modification parameter \(\alpha\) alters the transition behavior. In the case of large \(\ell\) and \(\alpha\), the transition occurs via a gradual decay of the QNM amplitude before shifting to the tail. In contrast, in other scenarios, the transition is more complex.

Lastly, a counterexample regarding the influence of \(\alpha\) on the tail amplitude is observed in the scenario shown in the upper-left corner of Fig.~\ref{2}. In this case, the amplitude increases with \(\alpha\), which is contrary to all other simulations. After thorough verification—including repeated calculations, adjustments to grid resolution, and modifications to time step sizes—we confirm that this result is not due to computational errors, but rather reflects a genuine numerical outcome. The interpretation of this phenomenon remains challenging and will be addressed in future work.

\medskip

\section{Discussion}

A key question arising from our study is whether the quantum corrections we observe could produce detectable signatures in the late-time tails of GWs. Within the framework of effective LQG, the dimensionless parameter \(\alpha\) introduced in this work is proportional to \(\sqrt{l_p / r_{BH}}\), where \(l_p\) is the Planck length and \(r_{BH}\) is the radius of the BH. This relationship suggests that larger values of \(\alpha\) can only emerge in extremely small BHs, such as those nearing the end of the BH evaporation process. Consequently, while these quantum effects could have an influence in such small-scale BHs, they are unlikely to impact typical astrophysical observations, where BHs are significantly larger.

It is important to note that the goal of this paper is not solely to compute quantum corrections to the late-time tails in BHs. Rather, the scenarios explored here, alongside those in \cite{Alnasheet:2025mtr}, should be viewed as part of a broader investigation into the potential deviations of real BHs from the predictions of GR. These deviations may arise due to unknown environmental effects around BHs in GR or from the influence of new physics, such as quantum gravity, on the structure of spacetime. From our results and those in \cite{Alnasheet:2025mtr}, it is clear that such deviations  affect the amplitude and transient behavior of the tails. These effects are not confined to a specific model; rather, they emerge in the results across different fields of study, each with distinct physically motivated background metrics (our work focuses on GWs, while \cite{Alnasheet:2025mtr} examines the behavior of a massless scalar field). Thus, the impact of quantum modifications or environmental influences on late-time tails is general and must be carefully considered when studying the intermediate behavior of tails, especially as we extrapolate these behaviors to QNMs. The fact that similar effects are observed across different models implies that the study of late-time tails could potentially be extended in a more model-independent way. It should a crucial development, especially in the context of future, more precise tests of GR and the nature of BHs. We will accomplish this in our future study.

In conclusion, although quantum corrections to the late-time tails may not be observable in current astrophysical systems, the investigation of these effects serves as an important step toward understanding potential deviations from GR in BH dynamics. The transition between QNMs and late-time tails must be approached with greater caution in future efforts to construct precise GW waveforms.

\medskip

\textit{Acknowledgements.}
QGH is supported by the grants from NSFC (Grant No.~12547110, 12475065, 12447101) and the China Manned Space Program with grant no. CMS-CSST-2025-A01.

\begin{appendices}
\section{Quick Review of NP Formalism}\label{NP}
In the NP formalism, we choose a pair of real-valued null vectors \( l^\mu \) and \( n^\mu \), along with a pair of complex-valued null vectors \( m^\mu \) and \( \bar{m}^\mu \), where \( m^\mu \) and \( \bar{m}^\mu \) are complex conjugates of each other. These vectors serve as the tetrad basis at each point of a four-dimensional pseudo-Riemannian manifold with signature \((-2)\) and metric \( g_{\mu \nu} \). The vectors satisfy the following relations:

\[
\begin{aligned}
l_\mu l^\mu = n_\mu n^\mu &= m_\mu m^\mu = \bar{m}_\mu \bar{m}^\mu = 0, \\
l_\mu n^\mu &= -m_\mu \bar{m}^\mu = 1, \\
l_\mu m^\mu = l_\mu \bar{m}^\mu &= n_\mu m^\mu = n_\mu \bar{m}^\mu = 0.
\end{aligned}
\]

Here, the notation \( \bar{m}^\mu \) indicates the complex conjugate of \( m^\mu \). The metric \( g_{\mu \nu} \) can then be expressed as:

\[
g_{\mu \nu} = l_\mu n_\nu + n_\mu l_\nu - m_\mu \bar{m}_\nu - \bar{m}_\mu m_\nu.
\]

The directional derivatives are defined as:

\[
D \equiv l^\mu \nabla_\mu, \quad \Delta \equiv n^\mu \nabla_\mu, \quad \delta \equiv m^\mu \nabla_\mu, \quad \bar{\delta} \equiv \bar{m}^\mu \nabla_\mu.
\]

Here, \( \nabla_\mu \) is the covariant derivative compatible with the spacetime metric \( g_{\mu \nu} \), such that \( \nabla_\mu g_{\nu \sigma} = 0 \).

These null vectors can be denoted as

\[
e_a^\mu = \left(l^\mu, n^\mu, m^\mu, \bar{m}^\mu\right) \quad (a = 1, 2, 3, 4).
\]

Consequently, the orthogonality condition can be expressed as \( g_{\mu \nu} = e_\mu^a e_\nu^b \eta_{ab} \), where

\[
\eta_{ab} = \eta^{ab} = \left( \begin{array}{cccc}
0 & 1 & 0 & 0 \\
1 & 0 & 0 & 0 \\
0 & 0 & 0 & -1 \\
0 & 0 & -1 & 0
\end{array} \right).
\]

Spin coefficients, also known as Ricci rotation coefficients, are defined as

\[
\gamma_{cab} \equiv e_c^\kappa e_b^\mu \nabla_\mu e_{a \kappa},
\]

where

\[
\eta_{ab} e_\mu^a = e_{b \mu}, \quad \eta^{ab} e_{a \mu} = e_\mu^b.
\]

Here is the definition of 12 complex spin coefficients in NP formalism:

$$
\begin{array}{lll}
\kappa=\gamma_{311} ; & \rho=\gamma_{314} ; & \varepsilon=\frac{1}{2}\left(\gamma_{211}+\gamma_{341}\right) ; \\
\sigma=\gamma_{313} ; & \mu=\gamma_{243} ; & \gamma=\frac{1}{2}\left(\gamma_{212}+\gamma_{342}\right) ; \\
\lambda=\gamma_{244} ; & \tau=\gamma_{312} ; & \alpha=\frac{1}{2}\left(\gamma_{214}+\gamma_{344}\right) ; \\
\nu=\gamma_{242} ; & \pi=\gamma_{241} ; & \beta=\frac{1}{2}\left(\gamma_{213}+\gamma_{343}\right) .
\end{array}
$$

According to the well-known decomposition of the Riemann tensor:
\[
\begin{aligned}
R_{abcd} =& C_{abcd} - \frac{1}{2} \left( \eta_{ac} R_{bd} - \eta_{bc} R_{ad} - \eta_{ad} R_{bc} + \eta_{bd} R_{ac} \right)\\
+& \frac{1}{6} \left( \eta_{ac} \eta_{bd} - \eta_{ad} \eta_{bc} \right) R,
\end{aligned}
\]
the curvature tensor can then be expressed in terms of the five complex Weyl scalars and ten NP Ricci scalars as follows:

\[
\begin{aligned}
\Psi_0 &= -C_{1313} = -C_{\alpha \beta \gamma \delta} l^\alpha m^\beta l^\gamma m^\delta, \\
\Psi_1 &= -C_{1213} = -C_{\alpha \beta \gamma \delta} l^\alpha n^\beta l^\gamma m^\delta, \\
\Psi_2 &= -C_{1342} = -C_{\alpha \beta \gamma \delta} l^\alpha m^\beta \bar{m}^\gamma n^\delta, \\
\Psi_3 &= -C_{1242} = -C_{\alpha \beta \gamma \delta} l^\alpha n^\beta \bar{m}^\gamma n^\delta, \\
\Psi_4 &= -C_{2424} = -C_{\alpha \beta \gamma \delta} n^\alpha \bar{m}^\beta n^\gamma \bar{m}^\delta.
\end{aligned}
\]

\[
\begin{aligned}
\Phi_{00} &= -\frac{1}{2} R_{11} = -\frac{1}{2} R_{\mu \nu} l^\mu l^\nu, \quad \Phi_{01} = -\frac{1}{2} R_{13} = -\frac{1}{2} R_{\mu \nu} l^\mu m^\nu, \\
\Phi_{10} &= -\frac{1}{2} R_{14} = -\frac{1}{2} R_{\mu \nu} l^\mu \bar{m}^\nu, \\
\Phi_{11} &= -\frac{1}{4} \left( R_{12} + R_{34} \right) = -\frac{1}{2} R_{\mu \nu} \left( l^\mu n^\nu + m^\mu \bar{m}^\nu \right), \\
\Phi_{02} &= -\frac{1}{2} R_{33} = -\frac{1}{2} R_{\mu \nu} m^\mu m^\nu, \quad \Phi_{12} = -\frac{1}{2} R_{23} = -\frac{1}{2} R_{\mu \nu} n^\mu m^\nu, \\
\Phi_{20} &= -\frac{1}{2} R_{44} = -\frac{1}{2} R_{\mu \nu} \bar{m}^\mu \bar{m}^\nu, \quad \Phi_{21} = -\frac{1}{2} R_{24} = -\frac{1}{2} R_{\mu \nu} n^\mu \bar{m}^\nu, \\
\Phi_{22} &= -\frac{1}{2} R_{22} = -\frac{1}{2} R_{\mu \nu} n^\mu n^\nu, \quad \Lambda = \frac{R}{24}.
\end{aligned}
\]
\end{appendices}

\bibliography{refs}
\end{document}